\documentclass[12pt,a4paper]{article}
\usepackage[british]{babel}

\usepackage[a4paper,top=2cm,bottom=2cm,left=2.5cm,right=2.5cm,marginparwidth=1.75cm]{geometry}
\usepackage{mystyle}

\begin{document}
\begin{center}
    {\LARGE \bfseries Application of Marchenko-based isolation to a land S-wave seismic dataset\par} 
    \vspace{0.5cm}
    \text{Running title: Marchenko-Based Isolation for Land Seismic Data}\par 
    \vspace{1cm}
    {\large Faezeh Shirmohammadi\textsuperscript{1}*, Deyan Draganov\textsuperscript{1}, Johno van IJsseldijk\textsuperscript{1}, Ranajit Ghose\textsuperscript{1}, Jan Thorbecke\textsuperscript{1}, Eric Verschuur\textsuperscript{1}, and Kees Wapenaar\textsuperscript{1}\par} 
    \vspace{0.5cm}
    \small
    \textsuperscript{1}Department of Geoscience and Engineering, Delft University of Technology, 2628 CN Delft, The Netherlands\par
    \vspace{0.5cm}
    \text{* corresponding author: f.shirmohammadi@tudelft.nl}\par
    \vspace{1cm}
    
\end{center}
\begin{abstract}
\noindent The overburden structures often can distort the responses of the target region in seismic data, especially in land datasets. Ideally, all effects of the overburden and underburden structures should be removed, leaving only the responses of the target region. This can be achieved using the Marchenko method. The Marchenko method is capable of estimating Green's functions between the surface of the Earth and arbitrary locations in the subsurface. These Green's functions can then be used to redatum wavefields to a level in the subsurface. As a result, the Marchenko method enables the isolation of the response of a specific layer or package of layers, free from the influence of the overburden and underburden. In this study, we apply the Marchenko-based isolation technique to land S-wave seismic data acquired in the Groningen province, the Netherlands. We apply the technique for combined removal of the overburden and underburden, which leaves the isolated response of the target region which is selected between 30 m and 270 m depth. Our results indicate that this approach enhances the resolution of reflection data. These enhanced reflections can be utilised for imaging and monitoring applications. 
\end{abstract}

\section{Introduction}
In seismic data, the focus is generally on target zones in the Earth’s subsurface, e.g. CO$_2$ and H$_2$ storage. However, the responses from these target zones are typically distorted due to interference from overburden and underburden structures, particularly in land seismic datasets. Therefore, it is essential to eliminate the effects of the overburden and underburden, which can be achieved using Marchenko method. The Marchenko method – a data-driven method – provides a tool for extracting information about the subsurface properties of the Earth. The Marchenko method retrieves Green’s functions in the subsurface from seismic reflection data at the surface. These Green’s functions can be used to redatum wavefields from the surface to arbitrary locations in the subsurface; a virtual source or receiver can be created at any point inside the medium of interest. This method employs reflection data from sources and receivers at the surface and an estimation of the first arrival, from a point in the subsurface from which we want to redatum to the surface. The first arrival can be estimated using a smooth velocity model (\cite{Slob2013SeismicEquations}; \cite{Wapenaar2014MarchenkoImaging}). \par
In recent years, there has been a significant progress in developing the Marchenko method and extending its applicability, for instance, for isolating the response of a specific subsurface layer without interference from the overburden and/or underburden. For an easier comparison with the original reflection data, the result of the Marchenko method can be extrapolated back to the surface (\cite{vanderNeut2016AdaptiveEquation}). This application results in a reflection response with sources and receivers at the surface and with fewer internal multiples from the overburden. This allows for more accurate characterisation of the properties of the target layer and can be particularly useful in target-oriented imaging and monitoring. \textcite{Wapenaar2021EMPLOYINGMETHOD} introduced the Marchenko-based isolation to identify the reservoir response from a seismic reflection survey by applying a two-step approach for removing the overburden and underburden interferences. \textcite{VanIjsseldijk2023ExtractingIsolationcite} showed that this application effectively isolates the target response, which can then be used to extract more precisely the local time-lapse changes in a reservoir. The Marchenko method for isolating a target response has been successfully applied to marine time-lapse datasets of the Troll Field for monitoring reservoir changes (\cite{IJsseldijkTrollb}). \par
As with any other method, the Marchenko method has some limitations. For its standard application, evanescent waves are ignored, the medium of interest is assumed to be lossless, and it is sensitive to time and amplitude inaccuracies in the reflection response. These facts can pose challenges, particularly when applied to noisy field data. Nevertheless, the method has been successfully applied to several marine field datasets for imaging and monitoring (\cite{Ravasi2016}; \cite{Jia2018}; \cite{Staring2018Source-receiverMethod}; \cite{Mildner2019}; \cite{Zhang2020AElimination}; \cite{IJsseldijkTrollb}). In spite of these advances, applications of the Marchenko method to land seismic data remain limited and more relatively problematic, not only because a reflection dataset free of surface waves and surface-related multiples is required, but also because the recorded data are inherently elastic (\cite{Reinicke2020}). \par
Here, we aim to apply the Marchenko-based isolation method to an SH-wave seismic dataset acquired close to the town of Scheemda in the Groningen province, the Netherlands, to isolate to isolate the target response, removing the overburden and underburden. The province of Groningen has been experiencing induced seismicity due to gas production since 1963 (\cite{Muntendam-Bos2022AnNetherlands}). Down to 800 m depth, the subsurface in Groningen comprises a sequence of soft, unconsolidated sediments, mainly composed of sand and clay (\cite{Kruiver2017}). Achieving an accurate depiction of the geometry and characteristics of these layers is crucial for precise earthquake studies. However, as known from Cone Penetration Testing (CPT) data, the first 30 meters of this site consist of alternating layers capable of generating strong internal multiples, which interfere with the response of all layers in the subsurface. Additionally, the deeper layers may generate arrivals that interfere with multiple reflections from the shallower layers. We acquired a reflection dataset specifically in this region to showcase the effectiveness of the Marchenko-based isolation method on land data and to enhance the visualisation of the reflection dataset in this region. \par
In the following section, we first review Marchenko-based isolation. Next, we describe the seismic acquisition parameters and the steps for preparing the input data for the Marchenko application. Finally, we discuss the results and how this study can contribute to future utilisation of the method, particularly for land-based applications. 

\section{Method}
Using the Marchenko method, we aim to isolate responses from a target layer by eliminating undesired events, including primaries and multiples, originating from both the overburden and underburden. To achieve this, the medium is divided into three units: overburden $\langle a \rangle$, target zone $\langle b \rangle$, and underburden $\langle c \rangle$, as shown in Figure~\ref{fig:6_1}. In the first step, the extrapolated Green's functions are determined with a focal level situated between the overburden $\langle a \rangle$ and the target zone $\langle b \rangle$ by using the Marchenko focusing functions. \par
Next, the overburden is removed, and the reflection response of the combined target zone $\langle b \rangle$ and underburden $\langle c \rangle$  is retrieved by applying seismic interferometry (SI) by multidimensional deconvolution (MDD, \cite{Broggini2014new}) with the extrapolated Green's functions. In the following step, the newly obtained reflection response is used to retrieve the extrapolated Marchenko focusing functions between the target zone $\langle b \rangle$ and the underburden $\langle c \rangle$. These retrieved Marchenko focusing functions are then used to remove the underburden using SI by MDD, effectively isolating the response of the target layer $\langle b \rangle$. \par
In the following subsections, we present the representations of Green's functions and then demonstrate how to isolate the responses of the target layer by eliminating the effects of the overburden and underburden.

\subsection{Green's functions representation}
The Marchenko method relies on two equations that relate the Green's functions and the focusing functions, derived from the one-way reciprocity theorems of the correlation and convolution types (\cite{Slob2013SeismicEquations}; \cite{wapenaarmarcehnko2014}; \cite{Wapenaar2021MarchenkoRelations}). Focusing functions are operators specifically designed to focus the wavefield, at a particular location within the subsurface. These functions are defined in a truncated medium, which matches the actual medium above the chosen focal level and is homogeneous below it. In the actual medium, focusing functions enable the wavefield to converge at the focal point, thereby facilitating the retrieval of a virtual source that generates Green's functions between the focal depth and the surface. The Marchenko method only requires the reflection responses at the surface and the first arrivals which can be estimated using a smooth velocity model of the subsurface.\par
\textcite{Meles2016} and \textcite{vanderNeut2016AdaptiveEquationcite} proposed extrapolating the virtual sources and receivers, obtained through Marchenko redatuming, back to the surface. This approach ensures that the travel times of events in the processed response remain consistent with those in the original reflection data, allowing for easier comparison of results before and after applying the Marchenko method. Both the focusing and the Green's functions are extrapolated using the direct arrival of the transmission response from the focal depth to the surface (\cite{vanderNeut2016AdaptiveEquation}), ensuring that the coordinates of all functions are located at the acquisition surface, \(S_{0}\). Following this approach, the coupled Marchenko extrapolated representations are defined as (\cite{Wapenaar2021MarchenkoRelations}; \cite{IJsseldijkTrollb}): 
\begin{equation}
   {U}^{-,+}(\mathbf{x}_{R},\mathbf{x}_{S}^{\prime},t) + v^{-}(\mathbf{x}_{R},\mathbf{x}_{S}^{\prime},t) = \int_{S_{0}} R(\mathbf{x}_{R},\mathbf{x}_{S},t) * v^{+}(\mathbf{x}_{S},\mathbf{x}_{S}^{\prime},t) d\mathbf{x}_{S},
   \label{eq:6_1}
\end{equation}

\begin{equation}
   {U}^{-,-}(\mathbf{x}_{R},\mathbf{x}_{S}^{\prime},-t) + v^{+}(\mathbf{x}_{R},\mathbf{x}_{S}^{\prime},t) = \int_{S_{0}} R(\mathbf{x}_{R},\mathbf{x}_{S},-t) * v^{-}(\mathbf{x}_{S},\mathbf{x}_{S}^{\prime},t) d\mathbf{x}_{S},
   \label{eq:6_2}
\end{equation}

\noindent which relate the extrapolated Green’s function \(U^{-,\pm}\) to the extrapolated focusing functions (\(v^{\pm}\)) using the reflection response \(R(\mathbf{x}_{R},\mathbf{x}_{S},t)\) at the acquisition surface (\(S_{0}\)). Here, \(\mathbf{x}_{R}\) and \(\mathbf{x}_{S}\) describe the receiver and source positions at the surface, respectively, and the superscripts $(-)$, $(\pm)$ of the extrapolated Green's functions represent an up-going receiver field from an up- $(-)$ or down-going $(+)$ source field, respectively. The asterisk ($*$) denotes temporal convolution. \par
Given that \(R\) is known, Equations~\ref{eq:6_1} and~\ref{eq:6_2} involve four unknowns:(\(U^{-,-}\), \(U^{-,+}\), \(v^{-}\), and \(v^{+}\)). To solve these equations, a window function is applied to suppress the extrapolated Green's functions. This approach relies on the fact that the extrapolated focusing functions and Green’s functions are separable in time (\cite{vanderNeut2016AdaptiveEquation}). This causality constraint can be applied after estimating the two-way travel time between the focal depth and the surface, which can be obtained from a smooth velocity model. By restricting Equations~\ref{eq:6_1} and~\ref{eq:6_2} to the calculated two-way travel time, the Green's functions on the left-hand side vanish. This simplifies the system to two equations with two unknowns (the extrapolated focusing functions), which can be solved iteratively (\cite{Thorbecke2017}) or through inversion (\cite{vander2015}). Once the focusing functions are found, the extrapolated Green's functions follow from Equations~\ref{eq:6_1} and~\ref{eq:6_2} without the time restriction. A detailed derivation of the Marchenko method is beyond the scope of this paper; however, a comprehensive derivation and background can be found in \textcite{Wapenaar2021MarchenkoRelations}.\par

\subsection{Over- and underburden removal}
\begin{figure}[h]
    \centering
    \includegraphics[width=\textwidth]{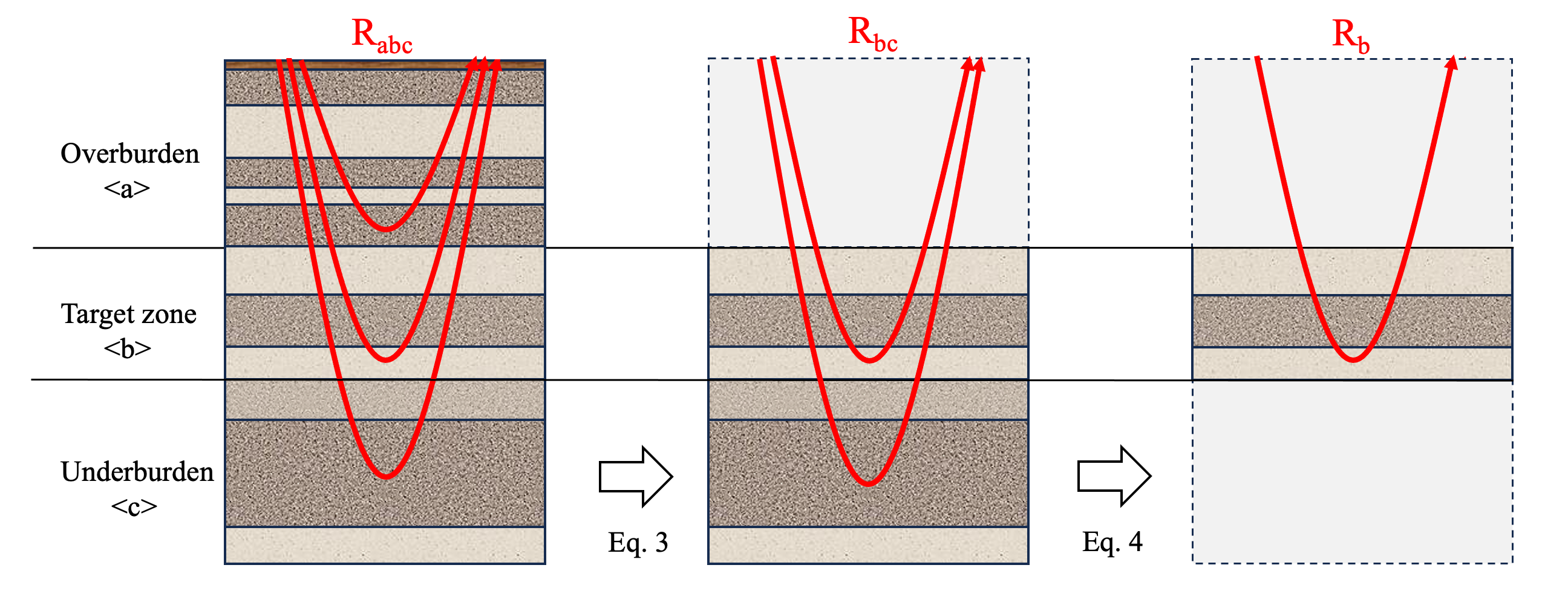} 
    \caption{Visual representation of the concept of Marchenko-based isolation. The medium is segmented into three units: overburden $\langle a \rangle$, target zone $\langle b \rangle$, and underburden $\langle c \rangle$. The initial step involves the removal of the overburden from the recorded response, as outlined in Equation~\ref{eq:6_3}. Following this, the response of the underburden is removed using Equation~\ref{eq:6_4} (adapted from \cite{IJsseldijkTrollb}).}
    \label{fig:6_1}
\end{figure}

The isolated responses of the target layer are retrieved through a two-step procedure. First, the overburden is removed, followed by the underburden removal. As introduced above, the medium is divided into three units: \(\langle a \rangle\) the overburden, \(\langle b \rangle\) the target layer, and \(\langle c \rangle\) the underburden, as shown in Figure~\ref{fig:6_1}. In the first step, for the overburden removal, a focal level is selected between the overburden \(\langle a \rangle\) and the target zone \(\langle b \rangle\), and Equations~\(\ref{eq:6_1}\) and~\(\ref{eq:6_2}\) are employed to determine the extrapolated Green's functions using the regular reflection responses (\(R_{abc}\)). For the overburden removal, the retrieved extrapolated Green's functions are employed to retrieve the reflection response isolated from the overburden interferences using:
\begin{equation}
    {U}_{a|bc}^{-,+}(\mathbf{x}_{R},\mathbf{x}_{S}^{\prime},t) = - \int_{S_{0}} {U}_{a|bc}^{-,-}(\mathbf{x}_{R},\mathbf{x}_{R}^{\prime},t) * R_{bc}(\mathbf{x}_{R}^{\prime},\mathbf{x}_{S}^{\prime},t) d\mathbf{x}_{R}^{\prime},
   \label{eq:6_3}
\end{equation}

\noindent where, \({{U}_{a|bc}^{-,\pm}}\) represents the extrapolated Green’s functions (\cite{Wapenaar2021MarchenkoRelations}, \cite{JohnoThesis}). The vertical line in the subscript indicates the location of the focal level, i.e., between the overburden $\langle a \rangle$ and the target zone and underburden $\langle bc \rangle$. Using Equation~\ref{eq:6_3}, the reflection response \(R_{bc}\) is retrieved employing SI by MDD. The retrieved \(R_{bc}\) contains all primary and multiple reflections from the target zone $\langle b \rangle$ and underburden $\langle c \rangle$ but it is devoid of overburden interactions from $\langle a \rangle$. Additionally, the coordinates $\mathbf{x}_{R}^{\prime}$ and $\mathbf{x}_{S}^{\prime}$ are situated at the surface, owing to the utilisation of the extrapolated Green’s functions (\cite{VanIjsseldijk2023aExtractingIsolation}). \par
For the next step, which is the underburden removal, this new reflection response (\(R_{bc}\)) can be utilised to retrieve the extrapolated focusing functions for a focal level between the target zone $\langle b \rangle$ and the underburden $\langle c \rangle$ using Equations~\ref{eq:6_1} and~\ref{eq:6_2} (\cite{WapenaarStaring2018}, \cite{IJsseldijkTrollb}). Then, to remove the underburden, we employ the following relation between the extrapolated focusing functions and the reflection response of the target (\(R_{b}\)):\par
\begin{equation}
    {v}_{b|c}^{-}(\mathbf{x}_{R},\mathbf{x}_{S}^{\prime},t) = \int_{S_{0}} {v}_{b|c}^{+}(\mathbf{x}_{R},\mathbf{x}_{R}^{\prime},t) * R_{b}(\mathbf{x}_{R}^{\prime},\mathbf{x}_{S}^{\prime},t) d\mathbf{x}_{R}^{\prime}.
   \label{eq:6_4}
\end{equation}
\par 
The subscript \(b|c\) indicates that the extrapolated focusing functions have been obtained from the reflection response without overburden interaction, utilising a focal depth between the target zone $\langle b \rangle$ and underburden $\langle c \rangle$. Once again, the isolated reflection response (\(R_{b}\)) can be retrieved from this equation through SI by MDD. Effectively, the target-zone response has now been isolated; the isolated response consists of consisting of the reflections (primaries and multiples) from inside it (\cite{IJsseldijkTrollb}). Figure~\ref{fig:6_1} shows this two-step procedure for removing the overburden and underburden using the Marcenko-based isolation.
\section{Seismic data acquisition}
\begin{figure}[H]
    \centering
    \includegraphics[width=\textwidth]{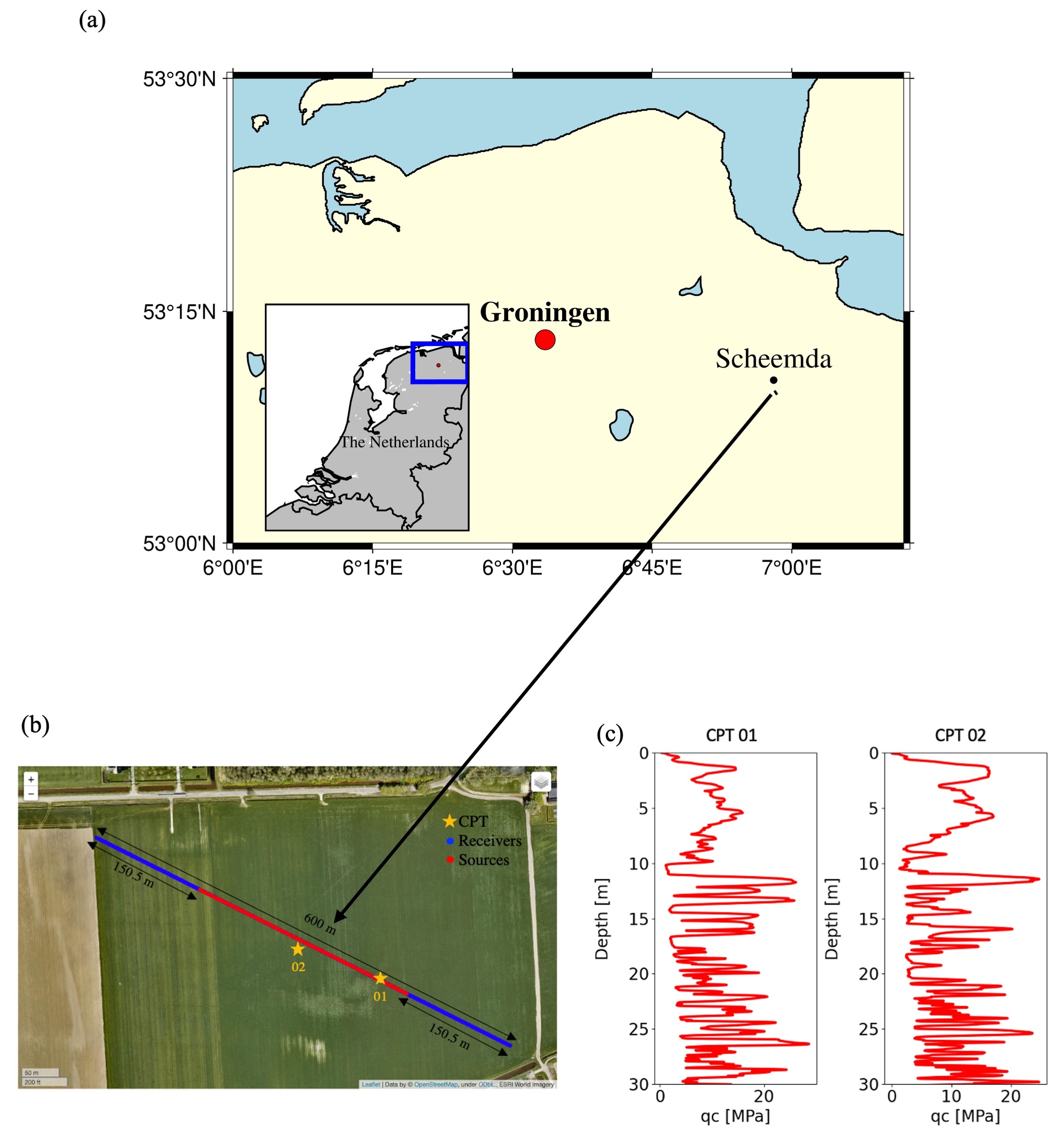} 
    \caption{(a) Location of the site, (b) the geometry of the reflection line. The red circles represent active sources, the blue circles represent receivers, and the orange stars represent the location of two cone penetration tests (CPT), and (c) cone-tip resistance (qc) measured at the two CPT location}
    \label{fig:6_2}
\end{figure}

In the summer of 2022, seismic reflection data was acquired along a line close to the town of Scheemda in the Groningen province of the Netherlands. Figures~\ref{fig:6_2}a and~\ref{fig:6_2}b show the location of the site and the geometry of the reflection line, respectively, and Figure~\ref{fig:6_2}c shows two examples of CPT data from this site. \par
We employed an electric seismic vibrator (\cite{Noorlandt2015AMotors}) as a source with a spacing of 2.0 m (the red circles in Figure~\ref{fig:6_2}b), and 601 three-component geophone nodes as receivers (the blue circles in Figure~\ref{fig:6_2}b), with a spacing of 1.0 m. The acquisition parameters are summarised in Table~\ref{tab:6_1}.\par
\begin{figure}[H]
    \centering
    \includegraphics[width=0.99\textwidth]{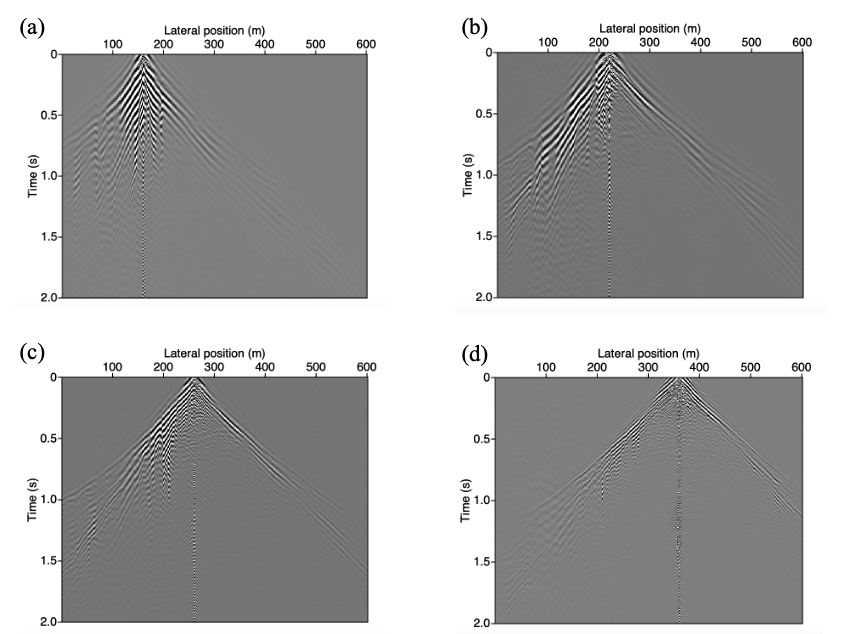} 
    \caption{Raw common-source gathers for active sources at lateral positions (a) 165.5 m, (b) 220.5 m, (c) 260.5 m, and (d) 360.5 m. }
    \label{fig:6_3}
\end{figure}

We made use of the vibrator in the S-wave mode oriented in the crossline direction. To apply the Marchenko method, we then use the data recorded by the crossline horizontal component of the geophones. Because of the orientation of the sources and the receivers, and assuming no scattering from the crossline direction, the SH-waves we record are generally decoupled from the compressional and vertically polarised S-waves. This makes the dataset more convenient for applying the Marchenko method. Four examples of common-source gathers at different source positions are shown in Figure~\ref{fig:6_3}.\par

\section{Data pre-processing}
The raw seismic reflection data cannot directly be used by the Marchenko method because when the method uses these data, it does not converge to a solution likely due to amplitude errors in the recorded data (\cite{Thorbecke2017}). As a first data pre-processing step, we apply source-signature deconvolution to obtain a high-resolution zero-phase reflection responses as input which is crucial for the Marchenko method. We use the designed source parameters as sent to the vibrator for a source-signature deconvolution. \par
In seismic land surveys with sources and receivers at the surface, the surface waves are most of the times dominant and they mask the reflections. This is also the case in our data. To eliminate surface waves, we apply surgical muting in the time domain and then employ bandpass filtering between 30 Hz and 100 Hz based on the power spectrum of the common-source gathers. To avoid any trace distortion, we chose not to use frequency-wavenumber filtering. \par 
The subsequent steps involved in preparing the reflection data for applying the \\ Marchenko method represent amplitude corrections. First, we correct the amplitudes for recording a 2D line in a 3D world. The effects of having geometric spreading in a 3D world while applying a 2D Marchenko scheme are corrected by applying a time-dependent gain to the data. This gain is approximately equal to $\sqrt{t}$ (\cite{Helegsen1993}; \cite{Brackenhoffpractical}). After that, we correct for absorption effects and for an overall amplitude mismatch that, for example, is related to the source signature. This is achieved by minimising a cost function as described in \textcite{Brackenhoff}. Different gains and linear factors are considered in order to find an optimal correction factor (\cite{JohnoThesis}). \par 
The Marchenko-method application uses the same number of sources and receivers, however, Our survey has more receivers positions than source positions. Therefore, we constrain our receivers to the extent and number of the sources, i.e., to the range 150 $-$ 450 m with 2.0 m spacing. \par
\begin{figure}[h]
    \centering
    \includegraphics[width=0.99\textwidth]{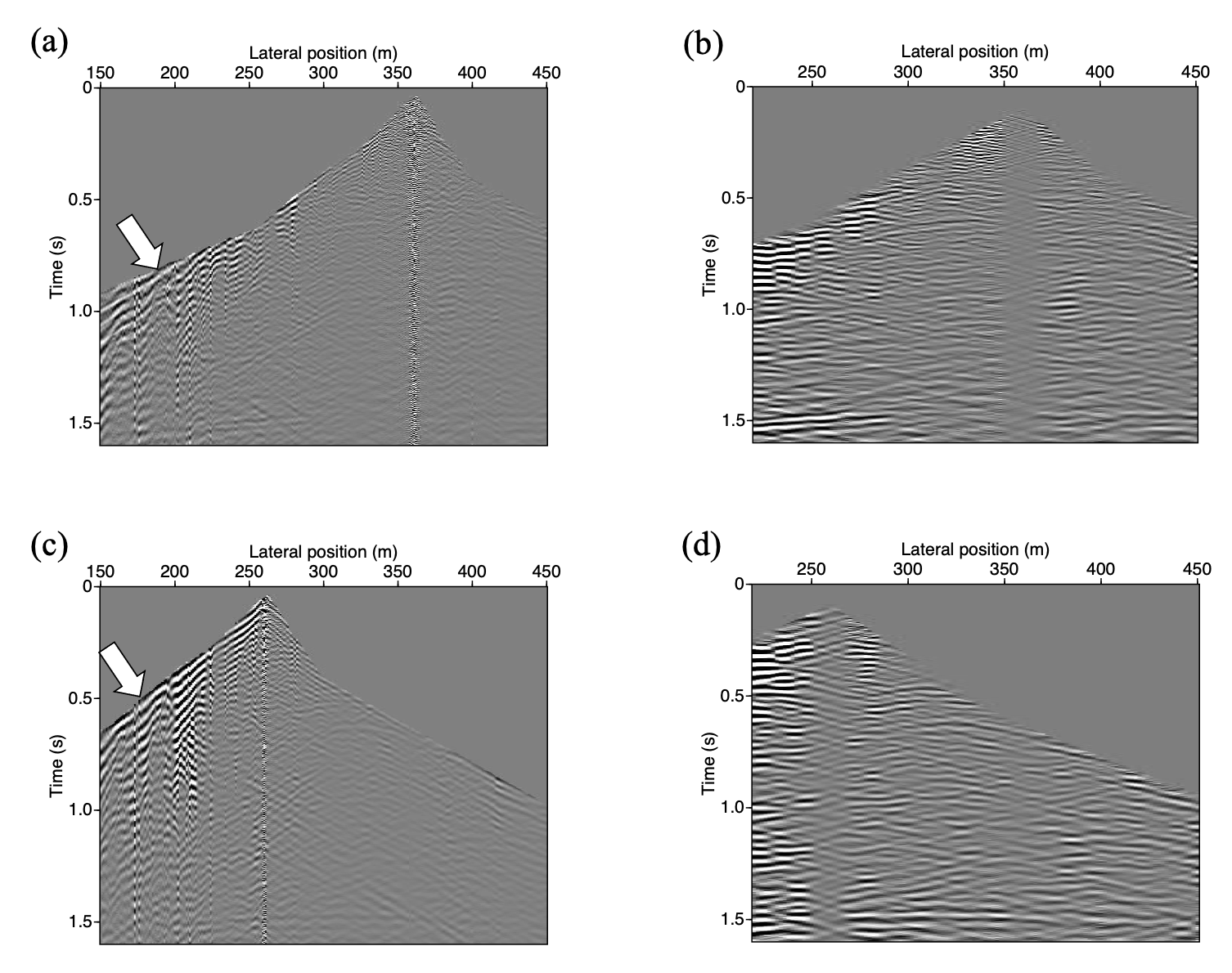} 
    \caption{Common-source gathers: (a) after the initial pre-processing steps and (b) after the final pre-processing steps for a source at a lateral position at 360.5 m. (c) and (d) are the same as (a) and (b), respectively, but for a source at 260.5 m. The white arrows show strong surface waves at the receiver positions at 150 $-$ 220 m.}
    \label{fig:6_4}
\end{figure}

Figures~\ref{fig:6_4}a and~\ref{fig:6_4}c show two common-source gathers after the application of the above-mentioned pre-processing steps. In these figures, it is evident that some pronounced surface waves persist at receivers positioned laterally between 150 m and 220 m, as indicated by the white arrows. Additionally, when comparing the common-source gathers shown in Figure~\ref{fig:6_3}, variations in the frequency content of the surface waves become apparent for sources located at lateral positions 150 m $-$ 220 m. This variation may occur due to local scattering in the shallow subsurface. Therefore, we further limit our sources and receivers to the range 220 m $-$ 450 m. Moreover, we see from the preprocessed gathers that the traces closest to the source location appear to have too strong amplitudes, i.e., they are strongly influenced by the source (near-field effects), which would result in difficulties during the inversion. Thus, we mute the 20 nearest traces around each source location, and then finally we apply a bandpass filter between 30 Hz and 100 Hz. Figures~\ref{fig:6_4}b and~\ref{fig:6_4}d show the common-source gathers from Figures~\ref{fig:6_4}a and~\ref{fig:6_4}c, respectively, after all pre-processing steps, showcasing enhancement in the visibility of the reflections. Additionally, we select only the earliest 1.6 s, as our primary focus lies in improving reflections up to this time. These processed data are then utilised as input for the Marchenko-based isolation technique.\par
\begin{figure}[h]
    \centering
     \includegraphics[width=0.3\textwidth]{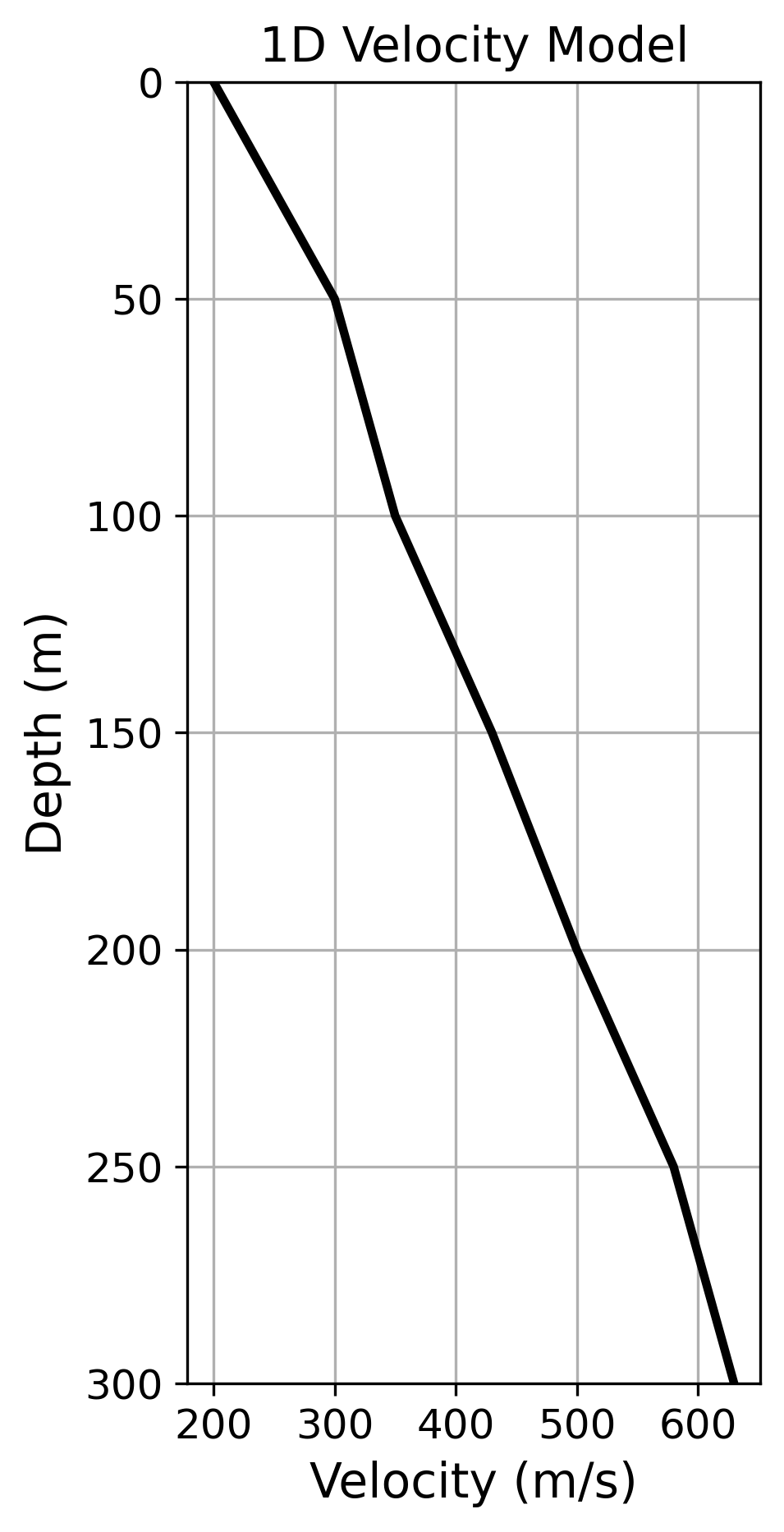}  
    \caption{A smooth 1D velocity model obtained from normal-moveout analysis on common-midpoint gathers.}
    \label{fig:6_5}
\end{figure}

As discussed in Section 2, we also require a smooth velocity model of the subsurface below the acquisition line to separate the focusing functions from the Green’s functions by estimating the two-way travel time between the surface and the focal depth. We obtain an 1-D smooth velocity model by standard Normal MoveOut (NMO) analysis applied to common-midpoint (CMP) gathers. Figure~\ref{fig:6_5} shows the estimated smooth 1D velocity model. 
\section{Results of Marchenko-based isolation}
\begin{figure}[h]
    \centering
    \includegraphics[width=0.99\textwidth]{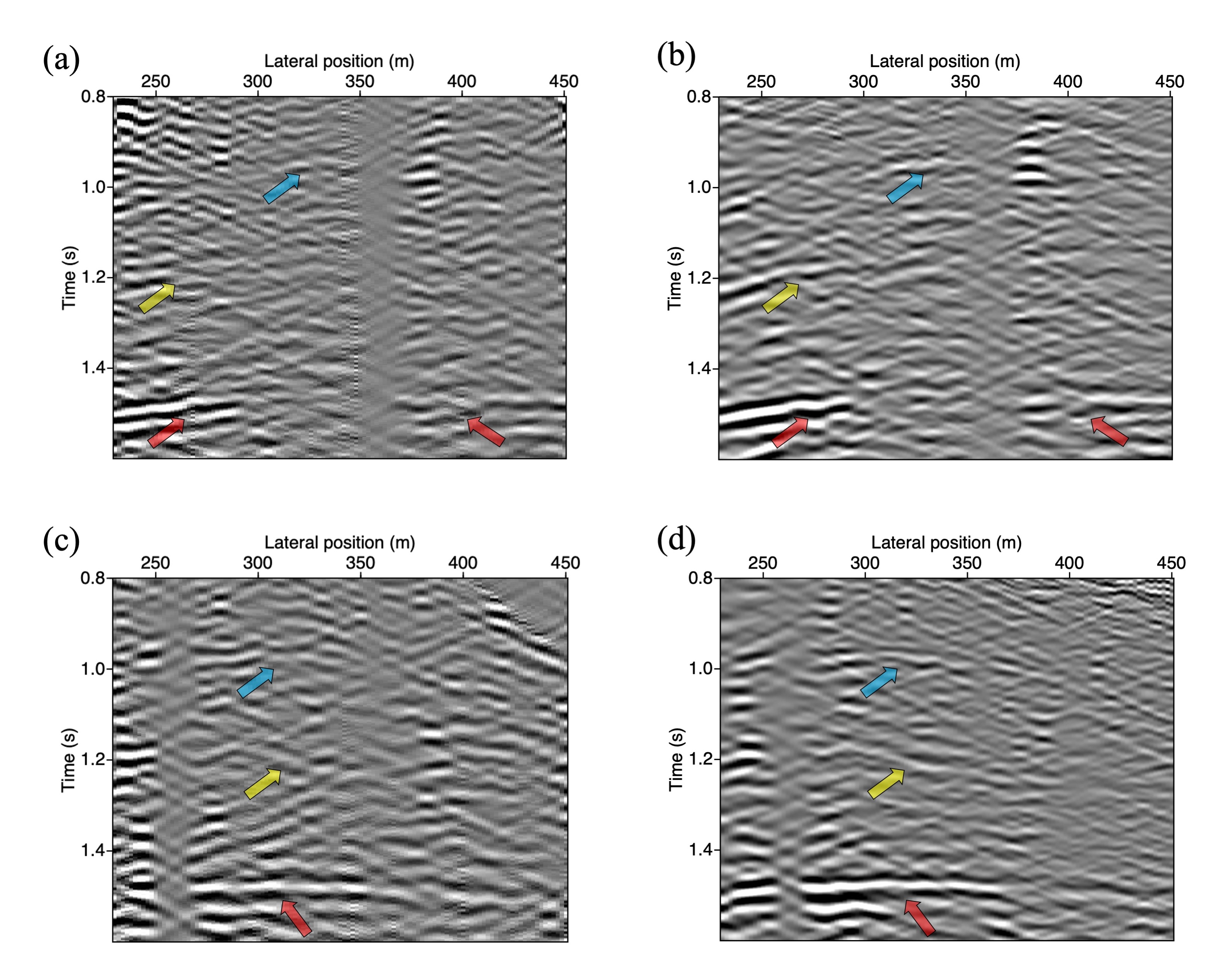} 
    \caption{A common-source gather for a source at a lateral position at 360.5 m (a) before and (b) after the Marchenko-based isolation. (c) and (d) are the same as (a) and (b), respectively, but for a source at 260.5 m. The cyan, yellow, and red arrows indicate reflections. The same bandpass filtering between 30 Hz and 100 Hz was applied for better comparison.}
    \label{fig:6_6}
\end{figure}

After completing the pre-processing steps, we employ the processed reflection data to extract the Green's functions and isolate the responses of the target region using the Marchenko-based isolation technique with one iteration. To account for both overburden and underburden effects, we employ the two-step procedure explained in Section 2. In the first step, we eliminate the overburden effect by choosing a focal depth of 30 m. Subsequently, using the results obtained from the first step, we eliminate the underburden effect by choosing a focal depth of 270 m. This two-step approach leaves the isolated response of the target region between 30 m and 270 m depth. Figure~\ref{fig:6_6} shows two common-source gathers before and after the Marchenko-based isolation. By comparing the common-source gathers before (Figures~\ref{fig:6_6}a and~\ref{fig:6_6}c) and after the Marchenko-based isolation (Figures~\ref{fig:6_6}b and~\ref{fig:6_6}d), it is evident that the resolution and clarity of some reflections are enhanced, as indicated by the cyan, yellow, and red arrows. \par
\begin{figure}[H]
    \centering
    \includegraphics[width=0.94\textwidth]{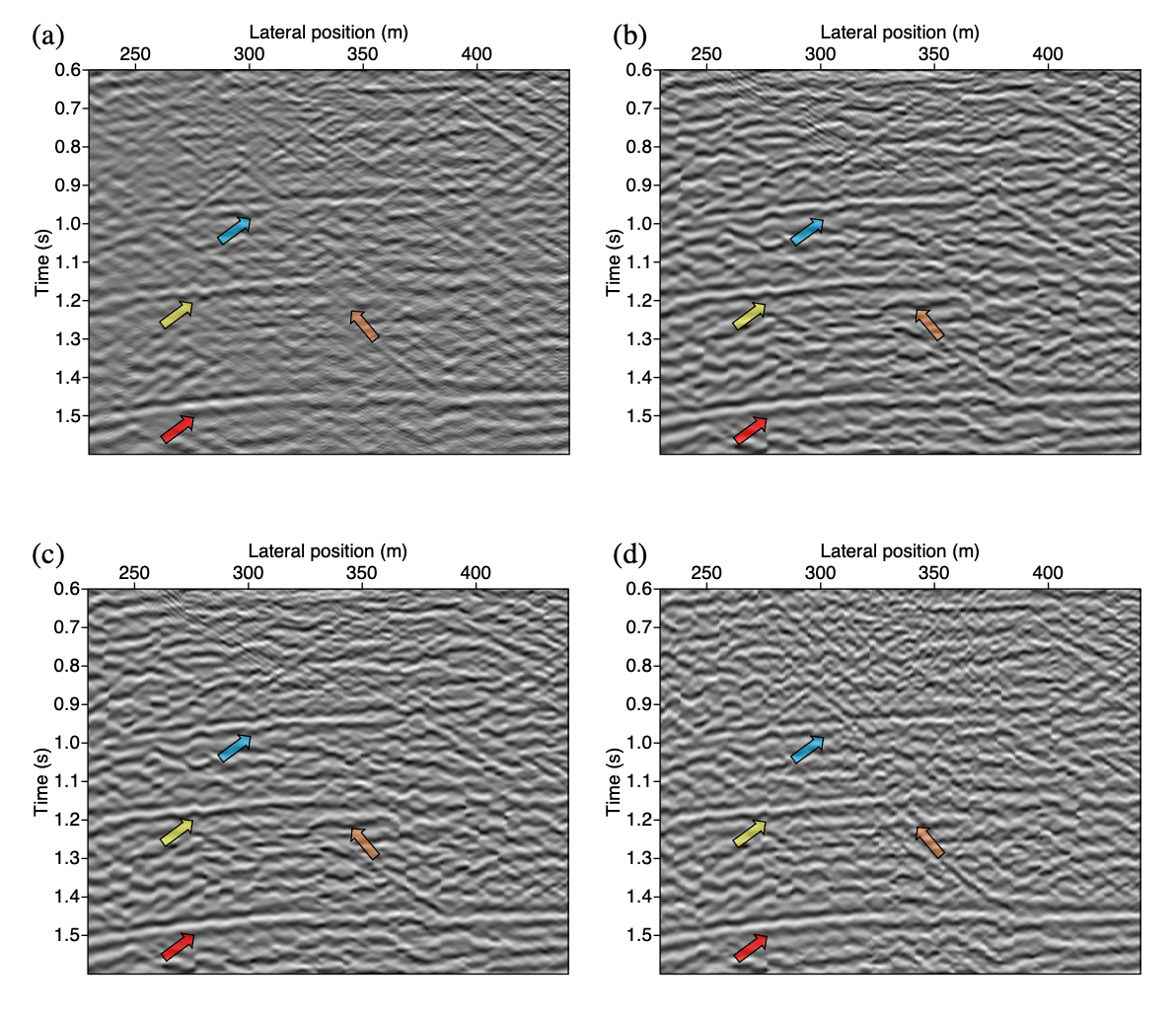} 
    \caption{The stacked sections obtained from (a) the regular reflection response and (b) the reflection response after the Marchenko-based isolation for overburden and underburden removal, (c) same as (b) but after only overburden removal, (d) same as (b) but after only underburden removal. The colour-coded arrows indicate reflections.}
    \label{fig:6_7}
\end{figure}

To facilitate a further comparison of the results, we perform NMO correction and stacking using CMP gathers for both the regular and the Marchenko-isolated responses. We use different constant velocities, an average velocity of the target region, and the time-varying NMO velocity for the NMO correction and stacking. By comparison, The constant velocity of 350 m/s leads to improved results regarding clear reflectors. We then apply an Automatic Gain Control (AGC) with time-window length 400 ms to aid the visual comparison. The stacked sections for both the regular and the Marchenko-based isolated responses are shown in Figure~\ref{fig:6_7}. \par
The stacked section obtained from the Marchenko-based isolated responses in Figure~\ref{fig:6_7}b is cleaner than the stacked section using the regular reflection response (with the same spatial extend) in Figure~\ref{fig:6_7}a. Moreover, the Marchenko-isolated stacked section exhibits more continuity, helping interpreting the data better. The events marked with cyan, yellow, orange, and red arrows indicate improvements in the stacked section in case of the Marchenko-based isolation. \par
To assess the individual effects of the overburden and underburden on the final result, we perform overburden and underburden removal separately. Figure~\ref{fig:6_7}c shows the stacked section after overburden removal, while Figure~\ref{fig:6_7}d shows the stacked section after underburden removal. By comparing Figure~\ref{fig:6_7}c and Figure~\ref{fig:6_7}d with Figure~\ref{fig:6_7}b, it becomes evident that most of the interaction originates from the overburden. The removal of the underburden does not improve the interpretability of the stacked sections, as enhancements are expected for arrivals after 1.42 s for a focal depth of 270 m using the estimated average velocity of 380 m/s from the smooth velocity model. This is likely due to small velocity variation in the underburden and attenuation of the reflected S-waves from the deeper layers. Additionally, the noise observed in Figure~\ref{fig:6_7}d may be attributed to the influence of the overburden. \par
\begin{figure}[H]
    \centering
    \includegraphics[width=\textwidth]{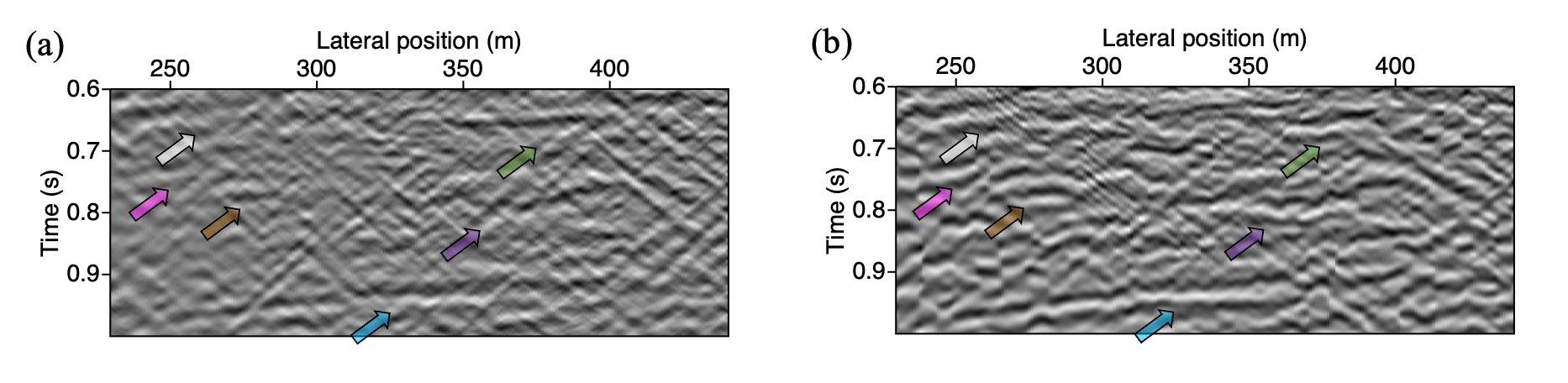} 
    \caption{Stacked sections, zoomed in between 0.6 s and 1.0 s, obtained using (a) the regular reflection response, (b) the reflection response after Marchenko-based isolation for overburden and underburden removal.}
    \label{fig:6_8}
\end{figure}

To better visualise the shallow part of the section, we further zoom in from 0.6 s to 1.0 s, see Figure~\ref{fig:6_8}. The geology in the top 30 m at this site, known from the cone-tip resistance (qc) measured at the two CPT location (Figure~\ref{fig:6_2}c), comprises of alternating clay and sand layers that contribute to the generation of internal multiples in this shallow section. Comparing the stacked section before Marchenko-based isolation in Figure~\ref{fig:6_8}a with the stacked section after Marchenko-based isolation in Figure~\ref{fig:6_8}b suggests a potential elimination of such internal multiples originating from the overburden down to 30 m in our case. Similar to the deeper reflectors, these shallow reflectors appear clearer and more continuous, as indicated by the colour-coded arrows. \par

\section{Discussion}
As demonstrated in the preceding section, the application of the Marchenko-based isolation to the target zone between 30 m and 270 m yields promising outcomes - it enhances  the quality of the reflection data and provides a clearer image of the target zone with more continuous reflectors. Even though we observe clear improvements in the reflection-data resolution and clarity, there are potential options for further enhancement in future studies and some extra measures to address the shortcomings of the Marchenko method. \par
First, the method is intentionally designed to minimise the reliance on a priori information, utilising only a smooth version of the velocity model and the recorded reflection responses. However, careful pre-processing is crucial to ensure a reflection response free from surface waves and surface-related multiples. In this study, we suppressed surface waves using surgical muting in the time domain and frequency filtering. It might be advantageous to eliminate all surface waves using techniques such as SI for surface-wave suppression (\cite{Liu2018SeismicWaves}; \cite{Balestrini2020ImprovedSuppression}; \cite{FaezehThesis}). \par
One of the constraints of the Marchenko method is its requirement for well-sampled and co-located sources and receivers. While our dataset was well-sampled, the sources were not precisely co-located with the receivers but were positioned 0.5 m away. However, this is acceptable given the expected minimum wavelength of 1 m, as the 0.5 m distance is still less than the minimum wavelength. For a dataset with irregular or imperfect sampling, \textcite{IJsseldijksampleing} demonstrated that the iterative Marchenko approach can be adapted to work with irregularly sampled data by using point-spread functions to recreate results as if they were acquired under perfect geometric conditions. \par 
One significant challenge in applying the Marchenko method to field data lies in the amplitude-scaling requirements on the reflection data. We tried to overcome amplitude mismatches by applying three gain factors for: geometrical spreading, absorption effects, and an overall scaling. The overall scaling factor has been investigated and tested for marine data (\cite{Brackenhoff}). As this is an important correction, it might be useful to investigate the overall scaling specifically for land seismic data and SH-waves.\par 
In this study, we applied the Marchenko-based isolation technique by selecting focal depths at 30 m and 270 m. We observed that the most significant contribution comes from the overburden removal. However, choosing different focal depths with smaller intervals and focusing on the improvement of specific reflections could be valuable in future studies.\par
Finally, the comparison of the common-source gathers and the unmigrated time sections, using a constant velocity for stacking, provided a good basis for our dataset, but we propose the application of a more realistic velocity analysis followed by migration. This additional step could contribute to a more comprehensive understanding and improvement of the Marchenko-based isolation technique.\par
\section{Conclusion}
We showed the result of applying the Marchenko-based isolation technique to SH-wave land seismic data that we acquired in the Groningen province, the Netherlands. Land seismic data known for dominant surface waves and low signal-to-noise ratio, Therefore, it is challenging to use the Marchenko method to land-seismic data, as the method ideally requires high-quality data. After careful implementation of surface-wave suppression, selection of the same spatial extent of sources and receivers, and the scaling-factor corrections as pre-processing steps, we retrieved the extrapolated Green’s functions, and the isolated target responses after the combined removal of the overburden and the underburden. \par
We showed that the resulting stacked section has improved signal-to-noise ratio, providing a better image of the target zone compared to the stacked regular reflection response. Our results open the door for future applications of the Marchenko method to land seismic datasets, particularly for time-lapse monitoring of the deeper structures, for example for CO$_2$ and H$_2$ storage, or for shallow applications such as monitoring waste management and changes in water table or other hydrological parameters.. \par

\section*{Data availability}
The field reflection dataset used in this study is available in the 4TU.ResearchData repository at \url{https://doi.org/10.4121/a8553b7e-82ae-4e9b-bc54-2a6b9ca6063c}. Codes associated with this study are available and can be accessed via the following URL: \url{https://gitlab.com/geophysicsdelft/OpenSource} in the “vmar” folder. More explanation can be found in \textcite{VanIjsseldijk2023aExtractingIsolation}. 

\section*{Acknowledgement}
We acknowledge the use of computational resources provided by the DelftBlue supercomputer at the Delft High Performance Computing Centre (\url{https://www.tudelft.nl/dhpc}) and thank Seismic Mechatronics for allowing us to use their vibrator source during fieldwork. (\url{https://seismic-mechatronics.com/}) \par
\noindent This research is funded by NWO Science domain (NWO-ENW), project DEEP.NL.2018.048 and the reflection dataset was acquired by funding from the European Research Council (ERC) under the European Union’s Horizon 2020 research and innovation programme (Grant Agreement No. 742703) and  OYO Corporation, Japan: OYO-TUD Research Collaboration (Project code C25B74).

\printbibliography[filter=exclude, heading=subbibintoc]
\newpage
\textbf{Tables}
\par
\begin{table}[htbp]
  \caption{Acquisition parameters}
  \small
  \centering
    \begin{tabular}{|c|c|}
     \hline
    \textbf{Parameter} & \textbf{Value} \\
    \hline
    Number of source positions & 151 \\
    \hline
    Source spacing  & 2.0 m  \\
    \hline
    First source position  & 150.5\\
    \hline
    Last source position   & 450.5 \\
    \hline
    Number of receiver positions per source & 601 \\
    \hline
    Receiver spacing & 1.0 m \\
    \hline
    First receiver position & 0 m \\
    \hline
    Frequency range of the sweep & 8-250 Hz \\
    \hline
  \end{tabular}
  \label{tab:6_1}
\end{table}

\newpage
\textbf{Figures}
\par

\newpage

\newpage
\newpage
\newpage
\newpage
\newpage
\newpage

\end{document}